\documentstyle[seceq,epsf,wrapft,twoside]{ptptex}
\notypesetlogo  
\title{From Yukawa to M-Theory
}
\author{%
Sho {\sc Tanaka}\footnote{Em. Professor of Kyoto University and Associate Member of Institute of Quantum Science, Nihon University. E-mail: stanaka@yukawa.kyoto-u.ac.jp}
}
\inst{%
Kurodani 33-4, Sakyo-ku, Kyoto 606-8331, Japan
}
\abst{%
Yukawa's space-time approach is briefly summarized, which starts
just before his meson theory and finally comes to the theory of 
elementary domain.  Recent rapid development in superstring theory 
is reviewed, focussing our attention on several important topics, 
which seem deeply related to Yukawa's concern. The importance 
of the idea of noncommutative space-time is emphasized.
}

\begin{document}
\maketitle

\setcounter{tocdepth}{4}
\section{Introduction}

Let me first refer to the historical background of my talk. I got the degree of doctor (thesis: ``Composite Model of Hadrons") under Professor S.~Sakata at Nagoya University in 1956, and  moved to Kyoto University at the end of 1957 after passing nearly two years at Rikkyo University, where I could come into contact with Professor M.~Taketani.
Professor Yukawa(1907-81) in Kyoto  was just 50 aged at that time. He was in the middle of research of nonlocal field theory. I remember that he was highly sensitive to the new proposal of Heisenberg's nonliner spinor theory(1953), which was regarded as a work aiming at ``a unified theory of universe". Surely, every things seemed exciting to us. Let me cite several important works in those days:
\begin{quote}
Heisenberg: S-Matrix (fundamental length $l_0$) (1943)\\
Yukawa: Nonlocal Field Theory (1947 - )\\ 
Snyder, Yang:  Quantized  Space-time  (1947)\\ 
Tomonaga, Schwinger, et al.: Renormalization Theory (1948)\\ 
Fermi-Yang:  Composite Model of $\pi$-meson  (1949)\\
Pauli-Villars:  Regularization  (1950)\\
Heisenberg:  Nonlinear Spinor Theory  (1953)\\
Nakano-Nishijima, Gell-Mann: Strange Particles (1953)\\
Yang-Mills:  Nonabelian Gauge Theory  (1954)\\
Sakata:  Composite Model of Hadrons  (1956)\\
\end{quote}

Here it is interesting to note that Yukawa's nonlocal field theory\cite{rf:nonl} and Snyder\cite{rf:snyder} -{Yang\cite{rf:yang}} 's quantized space-time theories appeared in the same year, 1947. Yukawa immediately noticed the Snyder's work, in the note added in proof of his paper. Indeed, the Snyder and Yang papers are the pioneer attempt of the present-day noncommutative space-time and will be often referred in the present talk.

I also remember very interestingly philosophical and methodological arguments among Sakata, Taketani and Yukawa about the different approaches in composite theory and nonlocal theory, as was seen later.   

\section{ Yukawa's Challenge to Conventional Space-time and \\ Local Quantum Field Theory (1934)}

  While Yukawa's full-scale investigation of nonlocal theory begins nearly at the end of 1940's, it should be noted, however, that its original idea goes back to April in 1934, that is, to our surprise, just in the midst of the proposal of his meson theory.  In the annual meeting of P-M Society, he talked ``On Probability Amplitude in Relativistic Quantum Mechanics"   (Yukawa, 1934)) in a strong sympathy with Dirac's similar idea of ``Generalized Transformation Function"  (Dirac, 1933).

Here, Dirac proposed generalization of concept of probability in quantum mechanics, by introducing integrals of lagrangian density over arbitrary form of space-time region, which was partially succeeded in Feynman's path integral (1946). 

Yukawa tried to find in this Dirac's idea a possibility of fundamental breakthrough to overcome the so-called 

``{\it Divergence Difficulty in Quantum Field Theory},"\\ 
which was explicitly pointed out by the famous paper of Heisenberg-Pauli (1929) and gave a serious influence on Yukawa over all his life to solve the problem. It ultimately led him to the firm belief:
\begin{quote}
{\it Fundamental particles must never be point-like, but have their own proper space-time extension.}  
\end{quote}

In fact, he tried to apply the above Dirac's action integral over arbitrary space-time region to the minimal space-time region of the order of Heisenberg's fundamental length $l_0$, by assuming that in such a minimal region, the conventional causal relation between definite initial and final times, $t_1$ and $t_2$ no longer holds, and {\it inseparability between cause and effect becomes essential.} Professor Sakata recollects\cite{rf:rec} the situation in the following expression that
\begin{quote}
 ``We the members of the physicist group in Japan can never forget that, at every conference in those days, he (Yukawa) used to draw a circle on the black board."
\end{quote}
Here, a circle, sometimes called  {\it Maru} in Japanese, implies the arbitrary space-time region mentioned above.
 
This Yukawa's proposition of idea of inseparability of cause-effect in the minimal region stimulated Professor Tomonaga to lead him to the so-called super-many-time theory (1943) and the renormalization theory of QED (1948), while Yukawa's idea of minimal region was there resolved in the form by replacing the region with one surrounded by the initial and final space-like hypersurfaces, i.e., ${\sigma}$-surfaces. Yukawa, of course,  could never be content with this solution to his aim. Indeed one finds that the proposition revives nearly 30 years later in his Theory of Elementary Domain (1966), as will be seen later.

\subsection{Bilocal Field Theory} 

On the challenge to the nonlocal field theory\cite{rf:nonl}, however, which begins in a full-scale in 1947, Yukawa seemed still careful to directly accept the above idea of minimal region, leaving space-time concept untouched, and attempted to introduce nonlocal field $U$ which is non-commutative with space-time coordinates
\begin{eqnarray}
[ U,  X_\mu ] \not= 0
\label{A}
\end{eqnarray}
in accord with Markov(1940).  Under the space-time coordinate representation basis, 
\begin{eqnarray}
| x_\mu \rangle,
\label{B}
\end{eqnarray}
which should be noted to be possible only for the case of commutative space-time, the above relation (\ref{A}) immediately leads us to the bilocal field:
\begin{eqnarray}
U ( x_\mu, x_\mu' ) \equiv \langle x_\mu| U | x_\mu' \rangle = U ( X_\mu , r_\mu)
\label{C}
\end{eqnarray}
In the above expression, one finds that bilocal field or two-point field $U (x_\mu , x_\mu')$ is rewritten in terms of center of coordinates  $X_\mu  = ( x_\mu + x_\mu' ) / 2$ and internal coordinates $ r_\mu = (x_\mu - x_\mu')$.

\subsection{Fierz's Decomposition}

As seen in the so-called Fierz's expansion or decomposition (1950),
\begin{eqnarray}
U ( X_\mu, r_\mu)  = \Sigma_k \int d\alpha\ \Phi_{\mu_1 \mu_2 \mu_3 \cdots \mu_k} ( X + \alpha r )  F_{\mu_1 \mu_2 \mu_3 \cdots \mu_k}(r)
\label{D}
\end{eqnarray}
one finds out that bilocal feld  turns out to be the infinite ensemble of local component fields $\Phi_{\mu_1 \mu_2 \mu_3 \cdots \mu_k} (X)$ with various spin $k$. 

This fact suggests that nonlocal field theory gives a possibility of ``Unified description of elementary particles". On the other hand, however, it leads us again to undesirable {\it local} component fields $\Phi_{\mu_1 \mu_2 \mu_3 \cdots \mu_k} (X)$, which have the danger of giving rise to the divergence difficulty. (Fierz, Hara-Shimazu(1950)).

Yukawa expected that this dilemma might be overcome by virtue of the nonlocal interactions of local component fields, whose nonlocal form-factor may suitably reflect the proper spatial extension of the original bilocal field, and provide cutoff of UV divergences. 

However, there remained unsolved many important questions like  causality or unitarity, other than the divergence problem, all of which, I think, have been left unsolved even in the present-day super-string theory, as will be remarked finally.

\subsection{From Bilocal to Multi-local Field}

Although Yukawa's bilocal field theory was formulated precisely and rigorously (1950) under the Lorentz-covariance, the reciprocity principle (Born, 1938), the correspondence principle to the local field theory and so on, the bilocal field itself seemed to be too limited to describe hadrons in general. Hence Yukawa and his collaborators tried to extend the bilocal field to multi-point fields:
\begin{eqnarray}
U (x_\mu, x_\mu') \rightarrow  U (x_\mu^{(1)},  x_\mu^{(2)} , x_\mu^{(3)} \cdots)
\label{E}
\end{eqnarray}

One sees, however, that this extension lacks a definite principle to formulate itself in contrast to bilocal field theory, and seems rather to imitate the wave-functions of many-body system in the composite model of hadrons.  

\subsection{ Urciton Scheme (Ishida, 1971)}

 In this connection, I would like here to notice Ishida's Urciton scheme (1971) as an attempt to connect both ideas of Yukawa's nonlocal theory and Sakata's composite theory, by introducing the exciton-like idea into hadronic level, which seems somewhat related to Yukawa's Elementary Domain or the present-day D(irechlet)-branes in superstring theories stated later.

\subsection{  String Model in the first stage or Pre-String Theory (1968-)}

The situation drastically changed, when the extreme limit of multi-point field was related to string field or string model of one-dimensionally extended object, $x_\mu (\sigma)$;
\begin{eqnarray}
U (x_\mu^{(1)},  x_\mu^{(2)} , x_\mu^{(3)} \cdots) \rightarrow U( x_\mu (\sigma) ),
\label{F}
\end{eqnarray}
which appeared immediately after Veneciano model (1968) devised  for the explanation of the dual behavior of Regge-poles and resonances in high-energy hadron reactions, and developed into the present-day superstring theory (Nambu, Susskind, 1968).

\subsection{ Elementary Domain (1966)}

 Before going into the problem, let me mention briefly on the Yukawa's idea of Elementary Domain\cite{rf:ED}. This was his first and final challenge to space-time structure itself, instead of a simple modification of point model of elementary particles as in nonlocal field.   As seen in the title of the first paper (1966), ``Atomistics and the Divisibility of Space and Time", it is clear that it originates in the idea of  "Minimal space-time region" proposed nearly 30 years ago, as was explained above.  

Yukawa tried to describe the minimal region, that is, the so-called elementary domain D, in terms of various parameters:

i)  Center of  Domain:
\begin{eqnarray}
X_\mu = \int_ D (d x)^4  x_\mu /  V_D
\end{eqnarray}

ii)  Moments of Extension:
\begin{eqnarray}
I_{\mu_1 \mu_2 \mu_3 \cdots \mu_n}  = \int_D (d x)^4 ( x_{\mu_1}  - X_{\mu_1}) \cdots ( x_{\mu_n} - X_{\mu_n}) / V_D,
\end{eqnarray}
with $V_D$ being the four-dimensional volume of D. They are to be observables describing various modes of deformation or excitation of an elementary domain, each of which are considered to correspond to the different kinds of elementary particles. 

Yukawa and his collaborators tried further to introduce Difference Equation in place of the usual differential equation, which gives the connection between the deformations of adjacent elementary domains. The theory, however, remained unaccomplished. Yukawa left us an impressive statement (1978) like: 
\begin{quote}
{\it If one proceeds along this way, it might ultimately lead to the problem of quantization of space-time itself. The concept of Elementary Domain may be insufficient, because of the fact that it still presumes behind it Minkowski-space or the four-dimensional continuum. The solution, however, must be left entirely in the future.}
\end{quote}

At this point, it is quite interesting to notice several important topics in the recent development of superstring theory, which seem deeply related to this Yukawa's concern or his long-sought goal. They are exemplified by the ideas of D(irichlet)-branes or holographic hypothesis, stated below.

\section{M-Theory}

As was already remarked, the original string theory appeared as hadron models at the end 1960's.  The situation drastically changed in the middle of 1970's, when it became clear that the closed string theory has a possibility of presenting naturally theoretical frameworks of quantum gravity as well as gauge theory. In fact, after the middle of 1980's, superstring theory became to be expected as a  unified theory of fundamental forces and matters including gravity. 
   During the past decade, especially after 1995, which is called the second stage of superstring theory, it occurred that the familiar five superstring theories tend to be unified, that is, TYPE-I, TYPE-IIA, IIB, and two Hetrotic String theories all formulated in $D=10$ are unified into a single fundamental theory, the so-called M-theory, F-theory, or S-theory in a hidden higher dimensional space-time, ($D$=11, 12, or 13). 

The situation is sometimes expressed on the analogy of the well-known historical experience of the unification of Schr{\" o}dinger's wave-mechanics and Heisenberg's  matrix-dynamics at the birth of quantum mechanics in 1920's, where it became clear that both approaches were merely due to the different choice of representation bases in Hilbert space.

As a matter of fact, the concept of string itself, as a linearly extended object, is not necessarily so drastic, but rather naturally conceivable in the continuous limit of multi-point particle system, as seen in (2.6). On the contrary, the so-called {\it Dirichlet-branes} seem more radical and remarkable, as will be shown below.

\subsection{Dp branes: p-dimensionally Extended Objects}

They were introduced (Polchinski\cite{rf:pol},1995) in the course of research of superstrings, and played an important role in unification of various kinds of superstring theories mentioned above.  In what follows, I would like to point out their important properties, keeping in mind the image sought in Yukawa's Elementary Domain.
\vspace{0.2cm}

( i ) \underbar{Dp Brane as Constraints on the End of Open String}
\vspace{0.2cm}

D(ichlet) p brane was originally introduced through the Direchlet boundary condition imposed on the open string, where the end points of open string are partially constrained on some $p$-dimensional hypersurface described with $p$ continuous variables, $\sigma_1,\sigma_2,\cdots, \sigma_p$ together with time parameter $\tau$.

This $p$-dimensional hypersurface is nothing but Dp brane. Under the suitable choice of coordinates in $D$-dimensional space, the space-time configuration of a Dp brane can be described in terms of $(D-p-1)$ coordinates :

\begin{eqnarray}
X_i (\tau,\sigma_1,\sigma_2,\cdots, \sigma_p ),\quad  i = p+1, p+2,\cdots, D-1
\end{eqnarray}
which are transverse to the remaining, i.e., the so-called $(p+1)$ world(-volume) coordinates, $x_\mu =  (\tau, \sigma_1, \sigma_2, \cdots, \sigma_p)$.
\vspace{0.2cm}

(ii) \underbar{ $U(1)$ gauge field  $A_\mu (x)$ living on a Dp brane} and

\qquad \underbar{Creation-Annihilation of Closed String} 
\vspace{0.2cm}

It is quite important to note that open string with both ends constrained on the same Dp brane produces $U(1)$-gauge field, $A_\mu(x)$ living on the Dp brane, which is one of the massless modes of the open string and regarded as a kind of local component fields, as was encountered in Fierz's decomposition of bilocal field, (2.4).  

In addition, it is noticeable that a Dp brane is concerned with creation and annihilation of closed string which is made through the joint of both ends of an open string on the Dp brane.

   These aspects lead us sometimes to the interesting view: {\it Dp brane is a Space-time Wall to the fundamental strings.} 
\vspace{0.2cm}

(iii) \underbar{ $U(N)$ Yang-Mills Gauge Fields living on $N$  Dp branes} and 

\qquad \underbar{Noncommutative Position Coordinates of Dp branes}
\vspace{0.2cm}

If there exist $N$  Dp branes, it becomes possible that open strings have their ends constrained on different Dp branes and each $U(1)$ gauge field on the respective Dp branes changes into $N \times N$ $U(N)$ Yang-Mills gauge fields  $\langle m|A_\mu (x)|n \rangle$ as a whole, when $N$ Dp branes precisely are on top of each other.

Furthermore, in this case, there arises a very important fact in describing the position coordinates of $N$ Dp branes:  Indeed, each Dp coordinate $X_i ( x_\mu )$ given in\\ ( i ), changes into $N \times N$  matrix form just like $U(N)$ gauge field $\langle m|A_\mu (x)|n \rangle$: 

\begin{eqnarray}
\langle m|X_i ( x_\mu)|n \rangle \qquad  m,n = 1,2,\cdots, N
\end{eqnarray}
Diagonal part, $\langle n|X_i ( x_\mu)|n \rangle$ may be naturally interpreted to describe 
the position coordinates of n-th Dp brane and nondiagonal part, $\langle m |X_i ( x_\mu) |n \rangle  (m \not= n)$ is explained due to the interaction between m-th and n-th Dp branes by open strings.

In this way, we encounter the idea of {\it the noncommutative position coordinates of Dp branes,} (Witten\cite{rf:witten}, 1995), that is, {\it entirely unexpected result from original string theories.}
\vspace{0.2cm}

(iv) \underbar{p-Brane Democracy and Superalgebra}
\vspace{0.2cm}

Dp brane so far explained might be seen still somewhat a secondary existence, in comparison with the fundamental strings. But it is quite important to see that their existence is well-founded in the frame work of superalgebra, such like 
$Osp(1\mid 32)$ in accordance with the idea of the so-called p-brane democracy (Townzend, 1995). 

One finds there that Dp branes participate in the formation of central charges as the essential ingredients, although we do not enter here into its detail. 

\subsection{D0 brane and Yukawa's Elementary Domain}

In the preceding arguments, (i)-(iv), I tried to summarize the remarkable aspects of Dp branes, with a certain expectation that the objects might be one of the promising candidates for the goal which Yukawa aimed through his idea of elementary domain, but did not achieve.  At this point, I should like to notice the paper\cite{rf:bfss}: `` M Theory as A Matrix Model: A Conjecture" (Banks, Fischler, Shenker and Susskind, 1996). In this paper, one finds that D0 brane with $p=0$, sometimes called D-particle, is regarded as the fundamental constituent of superstrings, and the supersymmetric matrix quantum mechanics of $N$-body system of D0 branes is proposed. 

At a glance, D0 brane seems to be a usual point-like particle, if we neglect the nondiagonal part of $X_i(\tau)$, which recalls to us the fundamental constituents which we have seen in the early string model as an infinite limit of multi-local field, as seen in section 2.5. However, one has to remember (i) that D0 brane is defined as a constraint on the ends of open string, and thus can never be separable from its background, just like the exciton or urciton discussed in section 2.4. Furthermore, one should remember that in the $N$-D0 brane system, the position coordinates as a whole must be described by $N \times N $ noncommutative matrices, as seen in (iii). 

These facts naturally lead us, further beyond the above idea of noncommutative position coordinates of D0 branes, to the idea of {\it noncommutative space-time}, that is, a definite departure from the continuous space-time itself and to the uncertainty in space-time domain, which were seriously sought in the Yukawa's Elementary Domain.

\subsection{Snyder-Yang's Quntized Space-time\cite{rf:snyder,rf:yang} (1947)}

At this point, one should recall the idea of Snyder-Yang's quantized space-time (1947), remarked in section 1. Needless to say, the Snyder-Yang's theories were proposed with the aim of solving ultraviolet divergence problem by virtue of the quantized or discrete space-time in place of the naive cutoff procedure. 

It is quite important to note that, among many recent proposals of noncommutative space-time algebra, Snyder-Yang's space is {\it discrete}, but {\it Lorentz-covariant}, as was emphasized by Yang (1965).

In addition, it should be pointed out that especially Yang's space-time algebra (YST) is deeply related to (Euclidean) conformal algebra (CFT) and to de Sitter algebra (dS), all of which have the common symmetry $SO(D+1,1)$. Indeed, the latter symmetry is noticed remarkably in the recent M-theory, the so-called dS/CFT correspondence or holographic hypothesis\footnote{At this point, it is very interesting to consider the modified version MYST of the original YST, by taking the background symmetry $SO(D+1,1)$ replaced with $SO(D,2)$, which is known to be common to $D$-dimensional conformal algebra and $(D+1)$-dimensional Anti-de Sitter algebra (AdS) and underlies AdS/CFT correspondence. Indeed, MYST is obtained from YST with two space-like extra dimensions $a$ and $b$ replaced with space-like $a$ and time-like $b$. See, hep-th/0303105\cite{rf:yqst}.}.

Taking into consideration these points, I attempted recently relativistic second quantization of the above quantum mechanics of D0 brane system\cite{rf:qst}, which makes creation and annihilation of D0 branes possible, by introducing second-quantized D0 brane field defined on the Yang's quantized space-time.

Here I have to omit its detail\cite{rf:yqst}, but briefly explain how the noncommutative $D$-dimensional space-time $X_\mu$ and momentum $P_\mu$ operators of Yang's quantized space-time algebra YST are defined through the dimensional contraction of $SO(D+1,1)$;
\begin{eqnarray}
X_\mu  = a\ {\Sigma}_{\mu a}
\end{eqnarray}
\begin{eqnarray}
P_\mu = 1/R\ {\Sigma}_{\mu b},
\end{eqnarray}
together with the reciprocal operator between $X_\mu$ and $P_\mu$,
\begin{eqnarray}
N= a/R\ \Sigma_{ab},
\end{eqnarray}
with fundamental constants, $a$ and $R$. In the above expression, ${\Sigma}_{MN}$ with $M,N = (\mu, a, b)$ is the angular momentum operator $SO(D+1,1)$ in the $D+2$ dimensional Yang's parameter space $\{ Y_M \}$ with two extra space-like dimensions, $a$ and $b$, and defined by
\begin{eqnarray}
{\Sigma}_{MN} = i\ ( Y_M \partial{}/ \partial{Y_N} -  Y_N \partial/ \partial{Y_M}).
\end{eqnarray}

\subsection{Holographic Bound, UV/IR Connection and Divergence Problem}

Now I would like to proceed to the final topics, the divergence problem, which was Yukawa's central concerns and has been discussed in the recent superstring theory in the form of UV/IR connection in holographic bound in AdS/CFT or dS/CFT correspondence. 

The idea of the holographic hypothesis is nicely expressed by Susskind-Witten (1998)\cite{rf:SW}: 
\begin{quote}
``According to the holographic hypothesis, a macroscopic region of space and everything inside it can be represented by a boundary theory living on the boundary of the region. Furthermore, the boundary theory should not contain more than one degree of freedoms per Planck area.  $\cdots$ One might imagine that the boundary theory is cutoff or discrete so that the information density is bound."
\end{quote}
\vspace{0.2cm}

In this argument, the following theories in the bulk and on the boundary, respectively, are nicely set up through the mediation of D3 branes (Maldacena, 1997); 
\begin{quote}
1) Type II B string theory with gravity on $AdS_5\times S^5$ for the bulk ($D=10$ macroscopic region) 

2)  $U(N)$ super Yang-Mills theory without gravity for the boundary, which is a conformal field theory.
\end{quote}
Conformal field theory on the boundary, however, is the conventional local field theory and contains ultraviolet divergence, which is connected with infrared divergence in the bulk AdS space with infinite boundary area in accordance with the so-called UV/IR connection. 

This fact naturally leads us to the above idea of Susskind-Witten, that is, a single cutoff parameter $\delta$ leads us in a unified way to the regularization of the infrared divergence in the bulk theory and the ultraviolet divergence in the boundary conformal theory. It provides the regularized Bekenstein-Hawking area-entropy relation, where the conformal field theory on the boundary contains one degree of freedoms per Planck area.
   
   At this point, one wonders naively why the ultraviolet divergence appears even in the superstring theory. It will be discussed in the last section.

\section{Concluding Remarks}

( I ) {\bf D0  branes and Yukawa's Elementary Domain}

I have pointed out several important results in the recent rapid development of superstring theory or M-theory, which seem to me deeply related to the fundamental problems of space-time. Especially I have noticed the concept of D0 branes as the fundamental constituent of superstring, in close accord with the idea of Elementary Domain, in which Yukawa dreamed the unified space-time description of elementary particles and the breakthrough to overcome the long-pending divergence problem.
\vspace{0.2cm}

(II) {\bf Divergence Problem in Superstring Theory}

On the other hand, however, we very often encounter various kinds of divergences in the effective low-energy theories of superstring theory, as seen in the UV/IR connection in the holographic bound, which may be related to local fields appearing as massless modes of superstrings.  At this point, we wonder why the ultraviolet divergences appear even in the superstring theory, while many authors assert that this problem already disappeared by virtue of nonlocality and supersymmetry intrinsic to the superstring theory. 

This fact recalls us the Yukawa's argument on the Fierz's local component fields in bilocal field theory in section 2.2, that is, the undesirable divergences accompanied with local component fields might be ultimately solved by taking into account the nonlocal form-factor of interactions which should suitably reflect the proper spatial extension of bilocal field. The naive cutoff or regularization procedure discussed in the UV/IR connection in section 3.4, for instance, might be understood as an effective substitute for such a nonlocal form-factor provided ultimately in superstring theory.
\vspace{0.2cm}

(III) {\bf Yang's Quantized Space-Time and UV/IR Connection}
 
 I would like, however, to emphasize more directly a possibility of the field theory on Yang's quantized space-time YST\cite{rf:qst} (or its modified version MYST\cite{rf:yqst}, see the footnote in section 3.3: (i) It has the common symmetry $SO(D+1,1)$ $(SO(D,2)$) underlying the dS/ CFT (AdS/CFT) correspondence. (ii) Discrete structure of its noncommutative space-time and momentum operators, which CFT lacks entirely, may provide a theoretical ground for the unified regularization or cutoff of UV-IR divergences, in place of the naive cutoff discussed in section 3.4.  

In order to arrive at the consistent divergence-free theory, however, many problems must be left in the future. 

\acknowledgements

The author would like to express his sincere gratitude to Prof. S.~Ishida and the research members of Nihon University for giving him a valuable chance to study the present big title and to talk in this symposium held by KEK and Nihon University, to which Professor Yukawa was deeply related.

\end{document}